\documentclass[runningheads]{llncs}
\usepackage[T1]{fontenc}
\usepackage{courier} 
\usepackage{listings, xcolor}
\usepackage{hyperref}
\usepackage{booktabs}
\usepackage{graphicx}
\usepackage{listings, xcolor}
\definecolor{javapurple}{rgb}{0.5,0,0.35} 
\definecolor{linenumbergray}{rgb}{0.5,0.5,0.5}
\lstdefinestyle{Java-github}{
        basicstyle=\ttfamily\tiny,
        language=Java,
        commentstyle=\color{linenumbergray},
        stringstyle=\color{javapurple},
        keywordstyle=\color{red},
        morekeywords={@Test},
        morecomment=[s][\color{linenumbergray}]{/**}{*/},
        numbers=left,
        numberstyle=\tiny\color{linenumbergray},
        numbersep=2.5pt,
        xleftmargin=1em,
        moredelim=**[is][\color{javapurple}]{@h@}{@h@},
        morecomment=[f][{\btHL[fill=gitdel]}]-,
        morecomment=[f][{\btHL[fill=gitadd]}]+,
        breaklines = true,
}
\lstdefinestyle{prompt}{
     basicstyle=\ttfamily\tiny,
     language=Html,
     commentstyle=\color{linenumbergray},
     stringstyle=\color{javapurple},
     keywordstyle=\color{red},
     morekeywords={@Test},
     morecomment=[s][\color{linenumbergray}]{/**}{*/},
     numbers=left,
     numberstyle=\tiny\color{linenumbergray},
     numbersep=2.5pt,
     xleftmargin=1em,
     moredelim=**[is][\color{javapurple}]{@h@}{@h@},
     morecomment=[f][{\btHL[fill=gitdel]}]-,
     morecomment=[f][{\btHL[fill=gitadd]}]+,
     breaklines = true, 
}
\begin{document}
\title{An approach for API synthesis using large language models}
%
%
\author{Hua Zhong \and
Shan Jiang \and
Sarfraz Khurshid}
\authorrunning{Hua et al.}
%
\institute{The University of Texas at Austin, Austin TX 78712, USA}
\maketitle              

\newcommand{\NumTasks}{135}
\newcommand{\NumTasksCompletedOurApproach}{133}

\begin{abstract}
APIs play a pivotal role in modern software development by enabling seamless communication and integration between various systems, applications, and services. Component-based API synthesis is a form of program synthesis that constructs an API by assembling predefined components from a library. Existing API synthesis techniques typically implement dedicated search strategies over bounded spaces of possible implementations, which can be very large and time consuming to explore.
In this paper, we present a novel approach of using large language models (LLMs) in API synthesis.  LLMs offer a foundational technology to capture developer insights and provide an ideal framework for enabling more effective API synthesis.  We perform an experimental evaluation of our approach using \NumTasks~real-world programming tasks, and compare it with FrAngel, a state-of-the-art API synthesis tool.  The experimental results show that our approach completes \NumTasksCompletedOurApproach~of the tasks, and overall outperforms FrAngel. We believe LLMs provide a very useful foundation for tackling the problem of API synthesis, in particular, and program synthesis, in general.



.
\end{abstract}

\section{Introduction}
The increasing complexity of modern software systems has made the correct usage of APIs and dependent libraries a significant challenge for developers\cite{lamothe2021systematic}. Jungloid mining \cite{jungloid} highlights that even experienced developers often spend substantial time identifying a few relevant functions within a library. Similarly, Shi et al. \cite{frangel} observed that programmers frequently explore libraries to locate useful functions, which often need to be combined with intricate control structures, such as loops and conditions, to implement desired functionalities. These challenges complicate program synthesis, as they require not only knowledge of the library's components but also the integration of those components into coherent, functional code.

API synthesis is particularly demanding because it requires an understanding of the intended behavior, which is usually only known to the developer and often lacks formal documentation suitable for automated reasoning. While various specification methods exist - such as logical constraints \cite{10.1145/1993316.1993506,proveri} and executable specifications \cite{mimic,oguide} - synthesizing APIs using input-output examples has emerged as a user-friendly alternative\cite{frangel}. This approach allows developers to communicate intended functionality through input/output examples for synthesis programs. This method is especially beneficial for non-programmers, who can use examples to articulate their requirements without needing technical expertise \cite{stringio}.

Previous approaches to component-based synthesis often rely on domain-specific knowledge to address particular problem classes, such as string manipulation \cite{icmlpbe,stringio} or data transformations \cite{syndstrans,syntable}. And test-driven-synthesis\cite{testdriven} describe a general framework that leverages expert-written domain-specific languages (DSLs). While other approaches, such as loop free synthesis\cite{loopfree} and oracle guided synthesis \cite{oguide}, are domain-agnostic but depend on formal component specifications. But such methods requires knowledge of expected behavior, which may only be known to the developer and may not not exist in a formal language that supports automated reasoning. However, writing formal specifications for every component in software libraries is labor-intensive and often impractical. While these synthesis approaches for complex APIs handle various practical synthesis problems, they are largely limited to creating single basic blocks of code and do not readily handle multiple blocks in the presence of loops (or recursion) and complex tests. A key issue with handling multiple blocks is the very large size of the space of possible method sequences and their combinations.

More recent methods, such as SyPet\cite{sypet}, EdSynth\cite{Edsynth}, and FrAngel\cite{frangel}, have aimed to reduce reliance on DSLs or formal specifications by using black-box execution instead of reasoning about formal semantics. While SyPet struggles with synthesizing programs that involve complex control structures like loops and conditionals, FrAngel introduces a strategy using angelic conditions and EdSynth eliminate this problem by sketching. FrAngel first identifies promising program structures and then resolves the control flow through optimistic evaluation. By iteratively refining control flows that pass test cases, FrAngel succeeds in combining known behaviors with control structures to uncover more complex functionalities \cite{pnondet}. LooPy\cite{loopy} makes use of live execution, a technique that leverages the programmer as an oracle to step over incomplete parts of the loop, which also got rid of the dependence on DSL. However, programmers as oracles also makes LooPy rely on the programmer and cannot automatically synthesis APIs.

In this paper, we propose an approach that leverage LLMs with prompt engineering for systematic study of synthesizing complex APIs. LLMs have recently shown exceptional capabilities in tasks involving both natural language understanding and code synthesis \cite{li2023cctest,nam2024using,jiang2024generating}. Trained on large datasets, including extensive code repositories and documentation, LLMs possess an implicit understanding of programming syntax and semantics. These models can align user inputs with their pre-existing knowledge \cite{liu2024llm} and even address tasks that are uncommon in their training data \cite{schaeffer2024emergent}. Unlike previous synthesis methods, which often rely on domain-specific knowledge or predefined formal specifications, LLMs excel at managing complex, open-ended programming tasks. They can synthesis APIs based on natural language prompts and simple input/output examples, eliminating the need for explicit specifications or DSLs. Furthermore, LLMs handle intricate control structures such as loops and conditionals with ease, making them highly versatile and efficient. By directly interpreting developer intent through input/output examples, LLMs reduce the burden on developers, enabling rapid and adaptive solutions for diverse functionalities.

An exciting consequence of using LLMs is that our approach requires minimal user input (e.g., fewer test cases than FrAngel) compared to other methods and provides richer implementations that include meaningful variable names and comments. The user just provides a few simplified input/output examples and the API signature and LLMs can synthesis correct API. In contrast, previous work like FrAngel have to define a complete set of component API which is the search space. The rich training corpus of LLMs allow them to be used without explicitly having to include all relevant technical details, such as complete lists of allowed APIs (components) to use. Checking program correctness -- a non-trivial task -- is simplified through the use of simple input/out examples, which allow developers to validate that the generated API meets their intended requirements. Developing compelete test cases remains a challenging problem and prior program-by-example (PBE) methods have struggled with corner cases. Our evaluation demonstrates that LLMs, due to their extensive training data and their ability to infer intended functionality from API signature, can handle corner cases with just a few test cases. Our approach simplifies the synthesis process while maintaining reliability and accuracy, offering a practical and accessible solution for developers. Specifically, our approach synthesizes 133 correct APIs out of 135 benchmarks, outperforms previous state-of-the-art technique, FrAngel, in accuracy and readability. 

In addition, to ensure that our method has generalization ability, we use the evaluation dataset in FrAngel which covers different complexity of APIs in different scenario. Besides, we design our additional dataset of 15 new APIs that are not present in open-source projects, which ensures that LLMs are truly capable of synthesizing API implementations based on input/output examples rather than simply reproducing content in their large training set.

This paper makes the following contributions:

\begin{itemize}
    \item {\bf Approach}.  We present a novel approach to use LLMs to do API synthesis.  By combining prompt techniques such as assistant context, chain-of-thought prompting, few-shot learning, and follow-up queries, our approach mitigates the need for exhaustive search and manual component specification.
    \item {\bf Evaluation}. We perform an experimental evaluation of our approach using \NumTasks~real-world programming tasks, and compare it to the state-of-the-art component-based techniques, FrAngel. Our experimental results indicate that our approach achieves over 98.5\% success rate in synthesizing correct, test-passing APIs, outperforming FrAngel in accuracy. Besides, our approach synthesizes readable APIs with meaningful variable names, comments, improving the clarity of the synthesized APIs.
    \item {\bf Artifact}. We release our prompt template in appendix and complete benchmark and experiment results at \href{https://github.com/shanjiang98/api-chatgpt}{https://github.com/shanjiang98/api-chatgpt}. This enables reproducibility of our results and offers a foundation for other researchers to explore LLM-powered program synthesis in additional contexts. 
\end{itemize}

\definecolor{javapurple}{rgb}{0.5,0,0.35} 
\definecolor{linenumbergray}{rgb}{0.5,0.5,0.5}
\lstdefinestyle{Java-github}{
        basicstyle=\ttfamily\tiny,
        language=Java,
        commentstyle=\color{linenumbergray},
        stringstyle=\color{javapurple},
        keywordstyle=\color{red},
        morekeywords={@Test},
        morecomment=[s][\color{linenumbergray}]{/**}{*/},
        numbers=left,
        numberstyle=\tiny\color{linenumbergray},
        numbersep=2.5pt,
        xleftmargin=1em,
        moredelim=**[is][\color{javapurple}]{@h@}{@h@},
        morecomment=[f][{\btHL[fill=gitdel]}]-,
        morecomment=[f][{\btHL[fill=gitadd]}]+,
        breaklines = true,
}

\section{Motivating Examples}
\label{sec:example}
This section walks through two example APIs, illustrates the limitations of existing Component-based API Synthesis techniques, and presents a novel API Synthesis technique that requires fewer user inputs and can generates more human readable implementations.

To illustrate the limitations inherent in current component-based API synthesis techniques, we walk through two examples from FrAngel's\cite{frangel} benchmark results. FrAngel is a state-of-the-art component-based API synthesis technique. In the first example, the task is to synthesize an API \texttt{getOffsetForLine} that takes a html document(Document) and a line(Integer) as the input arguments, and outputs the start offset(Integer) at index line of the document. This example was first studied in Sypet\cite{sypet}, another state-of-the-art component-based API synthesis technique, and FrAngel re-evaluated the Sypet dataset in its paper. The second example originates from the geometry benchmark of FrAngel's evaluation set. The target API, \texttt{ellipseArea}, calculates the area of an input ellipse. Fig.~\ref{fig:figure1} shows FrAngel's implementation of the two example APIs.

\begin{figure}
    \centering
    \begin{minipage}[t]{0.47\linewidth}
	\begin{lstlisting}[style = Java-github]
static int getOffsetForLine(Document doc, int line) {
    int var1 = 0;
    line = doc.getDefaultRootElement().getElement(var1).getEndOffset();
    return line;
}
        \end{lstlisting}
        \vspace{.5in}
        \begin{lstlisting}[style = Java-github]
static double ellipseArea(Ellipse2D ellipse) {
    return Math.PI * ellipse.getWidth() * ellipse.getHeight() / 4;
}
        \end{lstlisting}
    \vspace{.75in}
    \caption{\texttt{getOffsetForLine}(above) \& \texttt{ellipseArea}(below) APIs produced by FrAngel}
    \label{fig:figure1}
    \end{minipage}
\hspace{.2in}
    \begin{minipage}[t]{0.47\linewidth}
        \begin{lstlisting}[style = Java-github]
public int getOffsetForLine(Document doc, int line) throws BadLocationException {
    // Implement the logic to find the offset for the given line number
    Element root = doc.getDefaultRootElement();
    if (line < 0 || line >= root.getElementCount()) {
        throw new BadLocationException("Invalid line number", line);
    }
    Element lineElement = root.getElement(line);
    return lineElement.getStartOffset();
}
        \end{lstlisting}
	\begin{lstlisting}[style = Java-github]
public static double ellipseArea(Ellipse2D ellipse) {
    // The width and height of the Ellipse2D object represent the full lengths of the axes
    double semiMajorAxis = ellipse.getWidth() / 2;
    double semiMinorAxis = ellipse.getHeight() / 2;
        
    // The area of an ellipse is pi * semiMajorAxis * semiMinorAxis
    return Math.PI * semiMajorAxis * semiMinorAxis;
}
        \end{lstlisting}
        \vspace{.16in}
    \caption{\texttt{getOffsetForLine} (above) \& \texttt{ellipseArea} (below) APIs produced by the proposed methodology}
    \label{fig:figure2}
    \end{minipage}
\end{figure}

\noindent\textbf{User Input Test Cases.}
FrAngel is unable to produce an accurate solution in the first example because the original Sypet task provides an inadequate number of test cases. Most component-based synthesis methodologies employ constraint-based search approaches to generate code and they often rely on a set of comprehensive test cases to eliminate ambiguities in the task. However, comprehensive test case generation is non-trivial and challenging for complex APIs that encompass a substantial number of edge cases. Thus, existing techniques often require multiple input test cases to generate a solution.

\noindent\textbf{Code Readability.}
In the second example, the implementation generated by FrAngel presents challenges for human users to comprehend effectively. Existing techniques frequently employ non-descriptive variable names and omit intermediate steps during code generation. Additionally, they lack the ability to incorporate comments that would aid human programmers in understanding the resulting implementations.

These limitations underscore the necessity of developing a more advanced solution for complex API synthesis tasks. Our proposed approach, which leverages Large Language Models to generate the implementations, effectively mitigates these challenges and demonstrates promising results in our benchmark studies. As shown in Fig.~\ref{fig:figure2}, our approach successfully generates correct and human-comprehensible solutions for the two APIs, while requiring fewer user inputs.

\section{API Synthesis Methodology}
\label{sec:approach}
\begin{figure}[h]
    \centering
    \includegraphics[width=1.00\textwidth,trim={0cm 0cm 0cm 0cm},clip]{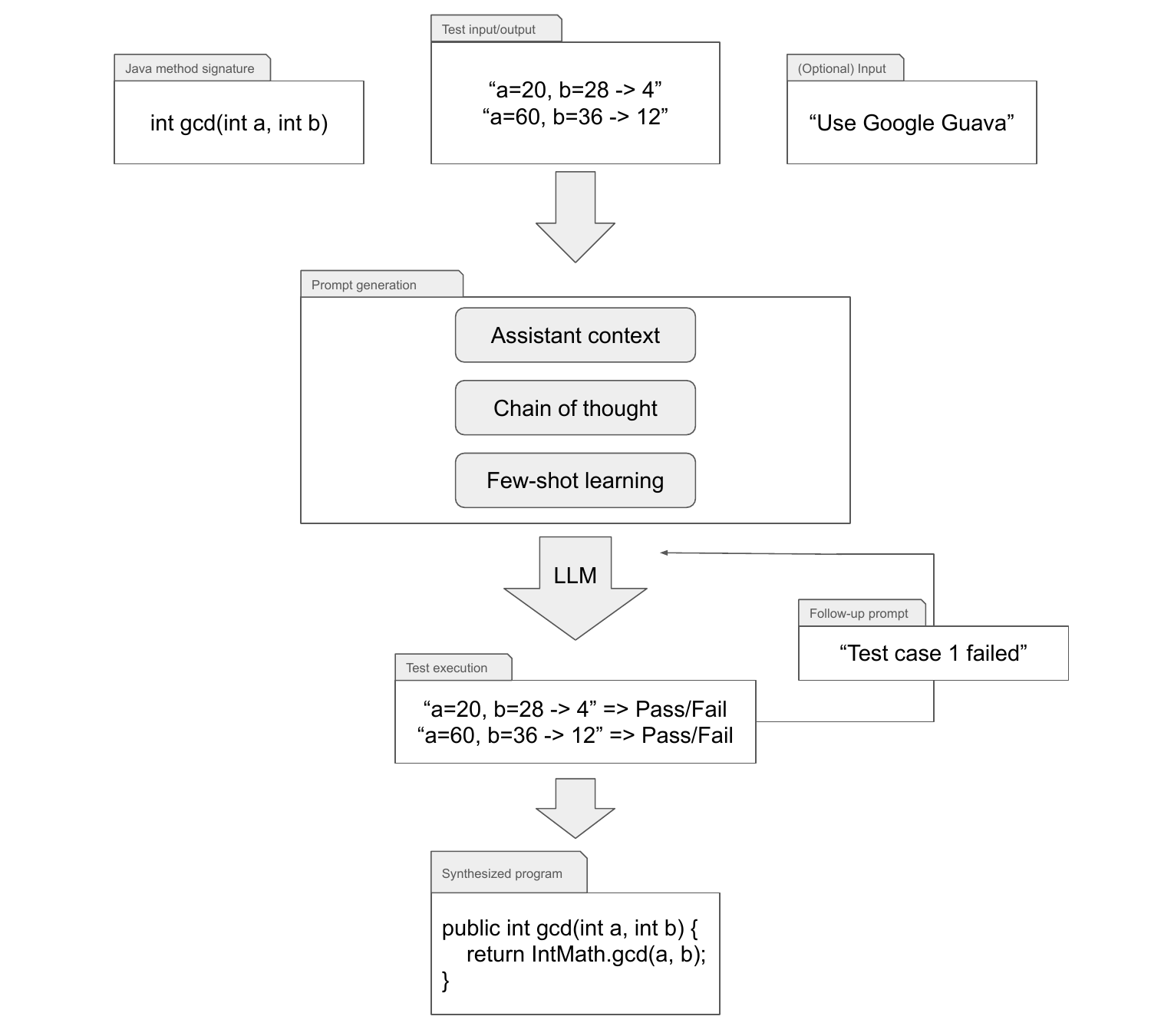}
    \caption{Workflow of the proposed API synthesis approach}
    \label{fig:figure3}
\end{figure}

In this section, we illustrate our novel approach on API synthesis, and break it down into different components. The overall workflow of our approach is shown in Fig.~\ref{fig:figure3}. Our study focuses on synthesizing Java APIs. In contrast to component-based methods that demand a list of libraries, our approach requires no input components, relying solely on a method signature and some test input/output pairs. We leverage the input to craft our API synthesis prompt and apply large language models (LLMs) to directly generate method implementations and test methods. We employ prompt engineering techniques to enhance the performance of LLMs and direct them to produce the intended output. The proposed methodology  can be broken down into several key steps: 
(1) Assistant creation which assigns a persona to the LLMs to strengthen its grasp of the task context;
(2) Chain of thought that divides a complex task into a series of simpler steps and produces intermediate outputs that collectively lead to the final result;
(3) Few-shot learning, a form of in-context learning, that enhances the performance of LLMs with a few examples of desired inputs and outputs to guide their responses;
and (4) Follow-up prompts provide feedback on the initial outputs of LLMs, aiding in the refinement and improvement of subsequent results. This process leverages the LLMs' capability for iterative enhancement and adaptation based on evaluative input.

We discuss the details of our prompt engineering technique in the following sections.

\subsection{Assistant Creation}
\label{sec:sec31}
Assistant creation is a widely adopted prompting technique that assigns a role to the LLM to guide it in solving a specific task. Assigning a role to an LLM establishes the problem context, enhancing its understanding of the task, and resulting in more accurate and relevant responses. As shown in Fig.~\ref{fig:figure4}, since the goal of our study is to synthesize a Java API from the method signature, we designate the role of LLMs to be a Java software engineer. This role assignment directs the LLMs to concentrate on Java programming-related reasoning, including the appropriate use of Java libraries, code readability, and adherence to best coding practices. In addition to role specification, the prompt explicitly conveys the task requirements to the LLMs as follows: "\textit{You will be given with an Java method signature, its return type, and a set of test cases as comments. Your task is to implement the Java method into a full implementation to pass the test cases.}" The prompt explicitly defines the task's input and expected output, effectively leveraging the programming knowledge of LLMs to facilitate API synthesis.

The process of assistant creation transforms LLMs from general-purpose models into domain-specific tools, optimizing their performance for specialized tasks. By establishing the assistant's role at the outset, we ensure that each subsequent phase such as parsing method signatures, implementing methods, and executing tests, is conducted with a targeted focus on Java programming principles and best practices.

\subsection{Chain of Thought}
\label{sec:sec32}
The chain-of-thought prompting strategy is a highly effective approach in prompt engineering, enabling LLMs to decompose complex tasks into a sequence of logically connected substeps. Our task of Java API synthesis presents a non-trivial challenge, necessitating logical reasoning to accurately interpret method signatures and their test cases. This process requires a structured sequence of intermediate steps to ensure a coherent and effective generation of APIs.

To develop an effective chain-of-thought (CoT) framework, we draw upon prior research\cite{wei2022chain} and begin by manually solving a representative subset of API synthesis tasks randomly selected from our benchmark dataset. Through this process, we identify and document the essential reasoning steps required for API synthesis. We then refine and optimize this reasoning procedure into an eight-step CoT framework. Finally, we leverage another LLM to further refine the manually generated steps, ensuring that the instructions are better aligned with the model's reasoning capabilities, thereby enhancing its effectiveness in solving the problem. \textit{Step 1} is designed to infer the necessary import statements and, most critically, to instruct the LLM to utilize relevant Java open-source libraries. This is a crucial aspect of legacy component-based API synthesis techniques, as it enables the integration of pre-existing components to facilitate efficient and accurate API generation. \textit{Step 2} directs the LLM to generate utility classes and/or methods to encapsulate auxiliary data structures, thereby supporting the implementation and enhancing modularity and maintainability. \textit{Steps 3} to \textit{Step 6} constitute the core stages that guide the LLM in method implementation. These steps instruct the LLM to define a class with a meaningful name that aligns with the method’s purpose, implement the method based on the provided signature, and ensure its correctness by considering all given test cases. Additionally, the LLM is directed to account for potential edge cases that are not explicitly covered in the provided test cases, thereby enhancing the robustness and reliability of the implementation. \textit{Steps 7} and \textit{Step 8} instruct the LLM to implement all provided test cases, ensuring comprehensive validation of the generated method. Additionally, the LLM is directed to execute the test cases and verify that the code compiles and runs without errors, thereby confirming the correctness and functional integrity of the implementation.

\subsection{Few-shot Learning}
\label{sec:sec33}
Instead of requiring training on extensive datasets comprising thousands or millions of examples, LLMs are known to have the ability to generalize to novel tasks or categories after being exposed to only a limited number of labeled examples\cite{fewshotahmed2023}.

To extract representative examples for few-shot learning, we manually selected a few examples from our benchmark system and then excluded them from the final evaluation set. Our selection criteria are as follows:
(1) To achieve a broader representation of various API types, we select five examples from different datasets within the benchmark system. 
(2) Examples containing an excessive number of tokens are filtered out to comply with the token limitations of LLMs.
(3) We ensure the inclusion of diverse method signature types, covering methods with and without return values, methods with primitive and non-primitive input parameters, and methods utilizing reference input parameters (i.e., methods that modify the referenced object).   
(4) Examples are selected to encompass varying numbers of associated test cases.
(5) We incorporate examples that both utilize and do not utilize library components to maintain a balanced representation.
Following this selection process, we identified five examples that satisfy the specified criteria, one of which is depicted in Fig.~\ref{fig:figure4}.

\subsection{Follow-up Prompt}
\label{sec:sec34}
In the context of LLMs, a "follow-up prompt" refers to a subsequent input provided after an initial prompt, enabling a more refined and context-aware interaction. This approach allows for iterative refinement by building upon prior responses, incorporating additional instructions, and steering the model toward a more precise or targeted outcome.
\begin{table}[h]
    \centering
    \begin{tabular}{p{0.22\linewidth} | p{0.73\linewidth}}
        \toprule
        \textbf{Error Type} & \textbf{Follow-up Prompt} \\
        \hline
        Compilation Error & The solution doesn't compile. Please fix the bug and make sure the implementation compiles, runs and can pass all test cases.\\
        \hline
        Run-Time Error & The solution throws run-time exceptions. Please fix the bug and make sure the implementation compiles, runs and can pass all test cases. \\
        \hline
        Failing Test Cases & The implementation fails test cases \textit{N} and \textit{M}. Please fix the bug and make sure the implementation compiles, runs and can pass all test cases.\\
        \bottomrule
    \end{tabular}
    \caption{Error type and follow-up prompt}
    \label{tab:table1}
\end{table}

In our approach, a follow-up prompt is issued to the LLMs when the initially generated API implementation is incorrect. Since errors can vary in nature, we employ distinct follow-up prompts tailored to address specific error types, ensuring a more effective refinement of the generated implementation. The various error types and their corresponding follow-up prompts are presented in Table~\ref{tab:table1}. If a follow-up solution fails with the same error as the previous attempt, the task is immediately classified as a failure. Therefore, up to three follow-up prompts are allowed to refine the generated API implementation before we fail the task.

\section{Evaluation}
\label{sec:evaluation}
In this section, we present a comprehensive experimental evaluation to assess the effectiveness of Large Language Models (LLMs) in Java API synthesis tasks. The primary objective is to evaluate the practicality of the generated APIs and their capability to accurately implement the expected functionality. Specifically, we address the following research questions:

\begin{itemize}
    \item \textbf{RQ1: Effectiveness.} What proportion of LLM-generated APIs successfully compile and pass the provided test cases?
    \item \textbf{RQ2: Comparison.} How does the success rate of our approach compare to traditional component-based API synthesis techniques?
    \item \textbf{RQ3: Correctness.} What is the frequency of incorrect solutions generated by our approach? 
    \item \textbf{RQ4: Code Readability.} How effective is our approach in generating code that is clear, readable, and easily understandable by humans?
\end{itemize}

\subsection{Experimental Setup}
\label{sec:sec41}
To address the above research questions, we designed an experimental framework incorporating the following key components:

\subsubsection{Large Language Models}
\label{sec:sec411}
For the LLMs, we choose the state-of-the-art GPT-4o model, a leading instruction-tuned language model, as the base model for our experiments. The temperature parameter is set to 0.7 to achieve a balance between creativity and consistency in the generated outputs. This configuration has been empirically demonstrated to be effective for code generation tasks\cite{endres2024can}. It enables the generation of diverse, yet controlled implementations. 

\subsubsection{Benchmarks}
\label{sec:sec412}
We evaluate our approach on five benchmark datasets, encompassing a total of 135 tasks. Specifically, we incorporate 90 tasks from the FrAngel benchmark and 30 tasks from the SyPet\cite{sypet} benchmark. Given that LLMs are likely trained on publicly available research papers and their associated benchmarks, we further introduce 15 novel tasks derived from various Java libraries to ensure the evaluation includes previously unseen scenarios, thereby assessing the generalizability of our approach. Furthermore, during the manual verification of the implementations, we observed that the solutions generated by LLMs differ significantly from those produced by previous approaches. This observation further reduces the likelihood that LLMs rely on having task solutions present in their training data to generate correct implementations.

\begin{table*}[h]
\centering
\caption{RQ1. Compilability (\textbf{Comp}) and execution success rate (\textbf{Pass}) with (\textbf{w/}) and without (\textbf{w/o}) follow-up prompts (\textbf{FuPs})}
\label{tab:table2}
\begin{tabular}{l|r|rr|rr}
\toprule
\multicolumn{1}{c|}{Benchmark} & \#Tasks & \begin{tabular}[c]{@{}c@{}}\#Comp w/o\\ FuPs\end{tabular} & \begin{tabular}[c]{@{}c@{}}\#Pass w/o\\ FuPs\end{tabular} & \begin{tabular}[c]{@{}c@{}}\#Comp w/\\ FuPs\end{tabular} & \begin{tabular}[c]{@{}c@{}}\#Pass w/\\ FuPs\end{tabular} \\ \hline
FrAngel & 90 & 88(97.8\%) & 83(92.2\%) & 90(100\%) & 88(97.8\%) \\
Sypet & 30 & 30(100\%) & 29(96.7\%) & 30(100\%) & 30(100\%) \\ 
Additional & 15 & 15(100.0\%) & 14(93.3\%) & 15(100.0\%) & 15(100.0\%) \\ \hline
Total & 135 & 133(98.5\%) & 126(93.3\%) & 135(100\%) & 133(98.5\%) \\
\bottomrule
\end{tabular}
\end{table*}

\subsection{Effectiveness}
\label{sec:sec42}
The first evaluation of LLM-generated APIs focuses on their effectiveness(\textbf{RQ1}), defined by the success rate of producing functional APIs. To assess this, we decompose the success rate into two key metrics: compilability and the ability to pass the provided test cases. We evaluate the compilability of the generated APIs in a standardized Java environment with JDK 21 and execute all provided test cases within the same environment.

Recall that in Section~\ref{sec:sec34} we discuss utilizing follow-up prompts to assist LLMs in correcting errors from their initial attempts. Consequently, we evaluate two sets of metrics to answer the research question. The details of these metrics are presented in Table~\ref{tab:table2}, which includes columns for the number of tasks and two sets of results: one for compilable tasks and test-passing tasks without follow-up prompts, and another for those incorporating follow-up prompts. We observe that over 98\% of the APIs generated by LLMs successfully compile without the assistance of follow-up prompts, and 100\% achieve compilability after the application of follow-up prompts. Among these, 93\% of the APIs successfully pass all test cases without the need for follow-up prompts. This success rate increases to over 98\% when follow-up prompts are utilized. Overall, our approach demonstrates high effectiveness in generating compilable APIs, achieving a 100\% compilation success rate. Furthermore, nearly all generated APIs (98.5\%) successfully satisfy the provided test cases.

The two unsuccessful tasks share a common root cause: the LLMs disregarded the prompt to utilize APIs from external libraries and instead attempted to implement complex logic independently. Fig.~\ref{fig:figure7} provides a detailed illustration of one such failure. Specifically, rather than invoking the \textit{JSONPath.eval()} API, the LLMs attempted to implement the traversal logic within the method, leading to the failure of test case \#5. It is important to note that this task could still be completed successfully by issuing additional follow-up prompts, either by explicitly specifying the \textit{JSONPath.eval()} API or by instructing the LLMs to correct the failure in test case \#5. However, doing so would violate the follow-up prompt strategy outlined in section~\ref{sec:sec34}. Consequently, we classify this task as a failure. Although the two tasks failed due to the inability to adopt suitable APIs, the LLMs successfully utilized appropriate APIs and/or implemented the necessary logic to pass all test cases in most other tasks.

\begin{figure}
\centering
    \begin{lstlisting}[style = Java-github]
public static Object fastjsonRead(String json, String path) {
    JSONObject jsonObject = JSON.parseObject(json);
        
    // Split the path by dots and brackets
    String[] pathParts = path.substring(2).split("\\[|\\]|\\.");

    // Traverse the JSON structure based on the path
    Object current = jsonObject;
    for (String part : pathParts) {
        if (part.isEmpty()) continue;
            
        if (current instanceof JSONObject) {
            current = ((JSONObject) current).get(part);
        } else if (current instanceof JSONArray) {
            JSONArray array = (JSONArray) current;
            if (part.contains("=")) {
                String[] filterParts = part.split("=");
                String filterKey = filterParts[0].trim();
                String filterValue = filterParts[1].trim();
                JSONArray filteredArray = new JSONArray();
                for (int i = 0; i < array.size(); i++) {
                    JSONObject obj = array.getJSONObject(i);
                    if (obj.get(filterKey).toString().equals(filterValue)) {
                        filteredArray.add(obj);
                    }
                }
                current = filteredArray;
            } else {
                int index = Integer.parseInt(part);
                current = array.get(index);
            }
        }
    }
        
    return current;
}

public static void main(String[] args) {
    // Test case 1 (PASS)
    JSONArray array1 = (JSONArray) fastjsonRead("{\"user\":[{\"amount\":1.11,\"isadmin\":true,\"age\":18},{\"amount\":0.22,\"isadmin\":false,\"age\":28}]}", "$.user[age = 18]");
    Assert.assertEquals(1, array1.size());
    Assert.assertTrue(1.11D == array1.getJSONObject(0).getDoubleValue("amount"));
    Assert.assertTrue(array1.getJSONObject(0).getBoolean("isadmin"));
    Assert.assertTrue(18 == array1.getJSONObject(0).getIntValue("age"));

    // Test case 2 (PASS)
    JSONArray array2 = (JSONArray) fastjsonRead("{\"user\":[{\"amount\":1.11,\"isadmin\":true,\"age\":18},{\"amount\":0.22,\"isadmin\":false,\"age\":28}]}", "$.user");
    Assert.assertEquals(2, array2.size());
    Assert.assertTrue(1.11D == array2.getJSONObject(0).getDoubleValue("amount"));
    Assert.assertTrue(array2.getJSONObject(0).getBoolean("isadmin"));
    Assert.assertTrue(18 == array2.getJSONObject(0).getIntValue("age"));
    Assert.assertTrue(0.22D == array2.getJSONObject(1).getDoubleValue("amount"));
    Assert.assertFalse(array2.getJSONObject(1).getBoolean("isadmin"));
    Assert.assertTrue(28 == array2.getJSONObject(1).getIntValue("age"));

    // Test case 3 (PASS)
    JSONArray array3 = (JSONArray) fastjsonRead("{\"user\":[{\"amount\":1.11,\"isadmin\":true,\"age\":18},{\"amount\":0.22,\"isadmin\":false,\"age\":28}]}", "$.user[isadmin = true]");
    Assert.assertEquals(1, array3.size());
    Assert.assertTrue(1.11D == array3.getJSONObject(0).getDoubleValue("amount"));
    Assert.assertTrue(array3.getJSONObject(0).getBoolean("isadmin"));
    Assert.assertTrue(18 == array3.getJSONObject(0).getIntValue("age"));

    // Test case 4 (PASS)
    JSONArray array4 = (JSONArray) fastjsonRead("{\"user\":[{\"amount\":1.11,\"isadmin\":true,\"age\":18},{\"amount\":0.22,\"isadmin\":false,\"age\":28}]}", "$.user[isadmin = false]");
    Assert.assertEquals(1, array4.size());
    Assert.assertTrue(0.22D == array4.getJSONObject(0).getDoubleValue("amount"));
    Assert.assertFalse(array4.getJSONObject(0).getBoolean("isadmin"));
    Assert.assertTrue(28 == array4.getJSONObject(0).getIntValue("age"));

    // Test case 5 (FAIL)
    JSONArray array5 = (JSONArray) fastjsonRead("{\"user\":[{\"amount\":1.11,\"isadmin\":true,\"age\":18},{\"amount\":0.22,\"isadmin\":false,\"age\":28}]}", "$.user[amount = 0.22]");
    Assert.assertEquals(1, array5.size());
    Assert.assertTrue(0.22D == array5.getJSONObject(0).getDoubleValue("amount"));
    Assert.assertFalse(array5.getJSONObject(0).getBoolean("isadmin"));
    Assert.assertTrue(28 == array5.getJSONObject(0).getIntValue("age"));
}
    \end{lstlisting}

\caption{One of the two programs generated by LLMs that failed to pass all test cases}
\label{fig:figure7} 
\end{figure}

\subsection{Comparison}
\label{sec:sec43}
We address \textbf{RQ2} by comparing the success rate of LLMs with a state-of-art component-based API synthesis technique, FrAngel. We execute all 120 tasks from the FrAngel's 4 benchmarks. The results are presented in Table~\ref{tab:table3}. Note that traditional component-based API synthesis techniques employ a slightly different strategy to determine task failure. Unlike our proposed approach, which assesses failure based on whether the generated code passes all provided test cases, traditional techniques rely on search algorithms and classify a task as a failure if the algorithm fails to identify a solution within a predefined time window. The time limit is set to 30 minutes for FrAngel. The LLMs utilized in our approach consistently generate responses within a few seconds, rendering the implementation of a timeout window unnecessary. Our approach slightly outperformed FrAngel across four benchmark datasets. A key insight from this comparison is that, unlike traditional techniques—which typically demonstrate strengths in a limited range of tasks—LLMs exhibit a high degree of adaptability across diverse task types. Additionally, as illustrated in Table~\ref{tab:table2}, the use of follow-up prompts substantially improves the success rate of LLMs. This iterative feedback loop plays a crucial role in guiding the LLM toward generating more accurate and contextually relevant responses. It is important to note that the tasks in which LLMs failed differ from those where previous approaches, such as FrAngel, encountered failures.

\begin{table*}[h]
\centering
\caption{Success rate of LLMs and FrAngel on four benchmarks.}
\label{tab:table3}
\begin{tabular}{l|r|r|r}
\toprule
\multicolumn{1}{l|}{Benchmark} & \#Tasks & \begin{tabular}[c]{@{}c@{}}FrAngel Success Rate \end{tabular} & \begin{tabular}[c]{@{}c@{}}LLMs Success Rate\end{tabular} \\ \hline
Geometry & 25 & 22(88.0\%) & 25(100.0\%) \\
ControlStructures & 40 & 38(93.3\%) & 40(100.0\%) \\
Github & 25 & 23(92.0\%) & 23(92.0\%) \\
Sypet & 30 & 30(100.0\%) & 30(100\%) \\ 
\hline
Total & 120 & 113(94.2\%) & 118(98.3\%) \\
\bottomrule
\end{tabular}
\end{table*}

\subsection{Correctness}
\label{sec:sec44}
To answer \textbf{RQ3}, we conducted a manual verification of a randomly selected subset of 43 programs from the 133 generated by LLMs. Additionally, we verified the 7 false positives reported in FrAngel's original study. A program is classified as a false positive if it successfully compiles and passes all test cases but fails to fully implement the intended functionality. Our analysis reveals that the LLMs successfully generated correct programs for all 50 tasks, including the seven tasks for which FrAngel failed to produce a correct solution. Fig.~\ref{fig:figure6} presents a side-by-side comparison of two programs generated by FrAngel and our LLM-based approach, highlighting the false positives produced by the former and the correct solutions generated by the latter.
\begin{figure}
\centering
    \begin{minipage}[t]{0.47\linewidth}
        \begin{lstlisting}[style = Java-github]
static double solution(Point2D point, Ellipse2D circle) {
    if (!circle.contains(point))
        point = new Point2D.Double(Double.MAX_VALUE, 0.0);
    return point.distance(circle.getCenterX(), circle.getCenterY());
}
        \end{lstlisting}
        \vspace{.82in}
        \begin{lstlisting}[style = Java-github]
static void guava_swap(Object[] array, int i, int j) {
    for (Object elem1 : array) {
        if (array[j] == elem1) {
            array[j] = array[i];
            array[i] = elem1;
        }
    }
}
        \end{lstlisting}
    \end{minipage}
\hspace{.2in}
    \begin{minipage}[t]{0.47\linewidth}
        \begin{lstlisting}[style = Java-github]
public double distanceInCircle(Point2D point, Ellipse2D circle) {
    double centerX = circle.getCenterX();
    double centerY = circle.getCenterY();
    double radiusX = circle.getWidth() / 2;
    double radiusY = circle.getHeight() / 2;
        
    double dx = (point.getX() - centerX) / radiusX;
    double dy = (point.getY() - centerY) / radiusY;
        
    double distanceSquared = dx * dx + dy * dy;
        
    if (distanceSquared <= 1) {
        return Math.sqrt(distanceSquared) * radiusX;
    } else {
        return Double.POSITIVE_INFINITY;
    }
}
        \end{lstlisting}
        \begin{lstlisting}[style = Java-github]
public static void guavaSwap(Object[] array, int i, int j) {
    if (array == null || i < 0 || j < 0 || i >= array.length || j >= array.length || i == j) {
        return;
    }
    Object temp = array[i];
    array[i] = array[j];
    array[j] = temp;
}
        \end{lstlisting}
    \end{minipage}
\caption{Two false positives produced by FrAngel(left), and the corresponding correct programs.}
\label{fig:figure6}
\end{figure}

False positives generated by traditional search-based techniques can typically be addressed through the inclusion of a comprehensive set of test cases. However, designing such a comprehensive test suite capable of eliminating all variations of false positives is a non-trivial task for developers, requiring substantial effort and domain knowledge. In contrast, LLMs possess the capability to comprehend the intended purpose of an API by leveraging method signatures and a few input/output examples, rather than merely generating solutions that satisfy the provided test cases.


\begin{figure}
	\centering
		\begin{lstlisting}[style = Java-github]
import java.util.LinkedList;
import java.util.Queue;
import java.util.Stack;
import java.util.List;

public class PalindromeChecker {

    boolean isPalindrome(Queue<Character> queue) {
        Stack<Character> stack = new Stack<>();
        Queue<Character> queueCopy = new LinkedList<>(queue);

        // Push all characters from the queue into the stack
        while (!queueCopy.isEmpty()) {
            stack.push(queueCopy.poll());
        }

        // Compare characters from the original queue and the stack
        for (Character ch : queue) {
            if (!ch.equals(stack.pop())) {
                return false;
            }
        }

        return true;
    }

    // Test methods for each test case
    public void testIsPalindrome() {
        // Test case #1
        Queue<Character> queue1 = new LinkedList<>();
        boolean result1 = isPalindrome(queue1);
        assert result1 == true : "Test case #1 failed";

        // Test case #2
        Queue<Character> queue2 = new LinkedList<>(List.of('a', 'c', 'c'));
        boolean result2 = isPalindrome(queue2);
        assert result2 == false : "Test case #2 failed";

        // Test case #3
        Queue<Character> queue3 = new LinkedList<>(List.of('r', 'b', 'g', 'b', 'r'));
        boolean result3 = isPalindrome(queue3);
        assert result3 == true : "Test case #3 failed";

        // Test case #4
        Queue<Character> queue4 = new LinkedList<>(List.of('c'));
        boolean result4 = isPalindrome(queue4);
        assert result4 == true : "Test case #4 failed";

        // Test case #5
        Queue<Character> queue5 = new LinkedList<>(List.of('c', 'c'));
        boolean result5 = isPalindrome(queue5);
        assert result5 == true : "Test case #5 failed";

        // Test case #6
        Queue<Character> queue6 = new LinkedList<>(List.of('b', 'c'));
        boolean result6 = isPalindrome(queue6);
        assert result6 == false : "Test case #6 failed";

        System.out.println("All test cases passed.");
    }

    public static void main(String[] args) {
        PalindromeChecker checker = new PalindromeChecker();
        checker.testIsPalindrome();
    }
}
            \end{lstlisting}

 \caption{A complete implementation of the \texttt{isPalindrome} API produced by LLMs, including the class definition, test methods, and the main method}
		\label{fig:figure5}
 
\end{figure}

\subsection{Code Readability}
\label{sec:sec46}
As discussed in Section~\ref{sec:example}, a key limitation of existing component-based API synthesis techniques is their inability to generate descriptive variable and method names. These approaches often overlook intermediate steps during code generation, leading to reduced clarity and maintainability. Furthermore, they lack the capability to incorporate meaningful comments, which are essential for enhancing the readability and comprehensibility of the generated implementations for human developers. Thus, to address the research question regarding code readability, we evaluate the generated APIs based on several key quality attributes:

\subsubsection{Naming and Comments}
\label{sec:sec461}
The Chain-of-Thought (CoT) section of the input prompts used in this study consistently instructs LLMs to generate functionally and semantically meaningful names for classes, methods, and variables, ensuring alignment with the problem statement and requirements of the API. For example, as illustrated in Fig.~\ref{fig:figure5}, the LLMs generate distinct and semantically meaningful names for variables, methods, and classes, thereby improving code clarity and maintainability. In the same example, the LLMs also demonstrate the ability to incorporate relevant comments into key statements or control flows of the program, enhancing code clarity and explicitly conveying its intended functionality.

\subsubsection{Code structure and clarity}
\label{sec:sec462}
The LLM-generated API methods show a clear and concise code structure, adhering to Java programming best practices and enhancing readability and maintainability. As shown in Fig.~\ref{fig:figure2}, the \textit{ellipseArea} method effectively employs local variables to document intermediate steps in the calculation of an ellipse's area. Additionally, the LLMs incorporate structured exception handling to manage potential runtime anomalies, enhancing the robustness and reliability of the implementation.

Overall, in response to \textbf{RQ4}, LLMs generate programs characterized by meaningful and expressive variable and method names, along with a clear and well-structured code organization, thereby enhancing their readability and maintainability for developers.


\section{Related Work}
\textbf{Program synthesis} has been a significant area of research, with various approaches proposed to generate programs efficiently. Traditional methods, such as Transit \cite{transit} and Escher \cite{escher}, rely on bottom-up enumerative synthesis, systematically exploring the program space to identify correct solutions. Brahma \cite{loopfree} introduces an efficient SMT-based encoding for synthesizing straight-line programs that involve multiple assignments to intermediate variables. Component-based synthesizers, such as SyPet \cite{sypet}, focus on generating Java programs from examples by leveraging arbitrary libraries. However, SyPet is limited to synthesizing sequences of method calls and does not support control structures like loops or conditionals. Similarly, Python-based synthesizers such as SnipPy \cite{snippy}, CodeHint \cite{codehint}, TFCoder \cite{tfcoder}, AutoPandas \cite{autopandas}, and Wrex \cite{wrex} primarily target one-liners or sequences of method calls. These tools also provide limited support for control structures, restricting their applicability for more complex program synthesis tasks. While these methods demonstrate the potential of automated program synthesis, their limitations in handling intricate control structures underline the need for more versatile approaches.

\textbf{Synthesis with control structures.} EdSynth \cite{Edsynth} address this by lazily initializing candidates during the execution of provided tests. The execution of partially completed candidates determines the generation of future candidates, making EdSynth particularly effective for synthesis tasks that involve multiple API sequences in both the conditions and bodies of loops or branches.
FrAngel \cite{frangel} is another notable tool that supports component-based synthesis for Java programs with control structures. Unlike many traditional methods, FrAngel relies on function-level specifications and, in principle, does not require users to have a detailed understanding of the algorithm or intermediate variables. However, in practice, to make the synthesis process feasible, users must provide a variety of examples, including base and corner cases. This requirement still necessitates some level of algorithmic knowledge, and FrAngel’s relatively slow performance makes it less suitable for interactive scenarios.
LooPy \cite{loopy}, introduces the concept of an Intermediate State Graph (ISG), which compactly represents a vast space of code snippets composed of multiple assignment statements and conditionals. By engaging the programmer as an oracle to address incomplete parts of the loop, LooPy achieves a balance between automation and interactivity. Its ability to solve a wide range of synthesis tasks at interactive speeds makes it a practical tool for use cases requiring real-time feedback and adjustments.
These tools demonstrate progress in addressing the challenges of synthesis with control structures, but they also highlight trade-offs between usability, required user input, and performance.

\textbf{Large Language Models (LLMs)} have recently shown effectiveness in various software development tasks, including program synthesis\cite{jain2022jigsaw} and test generation\cite{xia2024fuzz4all,jiang2024generating}. By associating document text with code from a large training set, LLMs can generate program code from natural language prompts\cite{jain2022jigsaw,ugare2024improving,spiess2024quality,wang2023large}. Reusable API\cite{reuseableAPI} uses LLMs to generate APIs from code snippets collected from Stack Overflow and shows significant results in identifying API parameters and return types. However, they provide all the code needed and only require LLMs to create new APIs using existing implementations. Test-driven program synthesis remains an under-researched topic. How well do LLMs perform in generating entire APIs with just a few input/output examples that even end users can easily prepare? In this paper, we explore the application of LLMs in generating API implementations and finds that LLMs are effective in understanding test cases and generating viable APIs. We also show that LLMs are able to generate accurate APIs even with an incomplete set of test cases, which is very convenient. To the best of our knowledge, our work is the first systematic study of using LLMs with prompt engineering to synthesis complex APIs.


\section{Conclusion}
In this paper, we present a novel approach of using large language models (LLMs) in API synthesis.  LLMs offer a foundational technology to capture developer insights and provide an ideal framework for enabling more effective API synthesis. The experimental results show that our approach overall outperforms FrAngel, a state-of-the-art API synthesis tool. Furthermore, our approach produces clear, maintainable code with meaningful variable names and relevant comments. Using minimal user output, our approach offers an easier way for API synthesis. We believe LLMs provide a very useful foundation for tackling the problem of complex API synthesis.

%
%

\bibliographystyle{splncs04}
\bibliography{main}

\begin{thebibliography}{10}
\providecommand{\url}[1]{\texttt{#1}}
\providecommand{\urlprefix}{URL }
\providecommand{\doi}[1]{https://doi.org/#1}

\bibitem{fewshotahmed2023}
Ahmed, T., Devanbu, P.: Few-shot training llms for project-specific code-summarization. In: Proceedings of the 37th IEEE/ACM International Conference on Automated Software Engineering. ASE '22, Association for Computing Machinery, New York, NY, USA (2023). \doi{10.1145/3551349.3559555}, \url{https://doi.org/10.1145/3551349.3559555}

\bibitem{escher}
Albarghouthi, A., Gulwani, S., Kincaid, Z.: Recursive program synthesis. In: Computer Aided Verification: 25th International Conference, CAV 2013, Saint Petersburg, Russia, July 13-19, 2013. Proceedings 25. pp. 934--950. Springer (2013)

\bibitem{autopandas}
Bavishi, R., Lemieux, C., Fox, R., Sen, K., Stoica, I.: Autopandas: neural-backed generators for program synthesis. Proc. ACM Program. Lang.  \textbf{3}(OOPSLA) (Oct 2019), \url{https://doi.org/10.1145/3360594}

\bibitem{pnondet}
Bodik, R., Chandra, S., Galenson, J., Kimelman, D., Tung, N., Barman, S., Rodarmor, C.: Programming with angelic nondeterminism. SIGPLAN Not.  \textbf{45}(1),  339–352 (Jan 2010). \doi{10.1145/1707801.1706339}, \url{https://doi.org/10.1145/1707801.1706339}

\bibitem{wrex}
Drosos, I., Barik, T., Guo, P.J., DeLine, R., Gulwani, S.: Wrex: A unified programming-by-example interaction for synthesizing readable code for data scientists. In: Proceedings of the 2020 CHI Conference on Human Factors in Computing Systems. p. 1–12. CHI '20, Association for Computing Machinery, New York, NY, USA (2020). \doi{10.1145/3313831.3376442}

\bibitem{endres2024can}
Endres, M., Fakhoury, S., Chakraborty, S., Lahiri, S.K.: Can large language models transform natural language intent into formal method postconditions? Proceedings of the ACM on Software Engineering  \textbf{1}(FSE),  1889--1912 (2024)

\bibitem{syntable}
Feng, Y., Martins, R., Van~Geffen, J., Dillig, I., Chaudhuri, S.: Component-based synthesis of table consolidation and transformation tasks from examples. In: Proceedings of the 38th ACM SIGPLAN Conference on Programming Language Design and Implementation. p. 422–436. PLDI 2017, Association for Computing Machinery, New York, NY, USA (2017). \doi{10.1145/3062341.3062351}, \url{https://doi.org/10.1145/3062341.3062351}

\bibitem{sypet}
Feng, Y., Martins, R., Wang, Y., Dillig, I., Reps, T.W.: Component-based synthesis for complex apis. In: Proceedings of the 44th ACM SIGPLAN Symposium on Principles of Programming Languages. p. 599–612. POPL '17, Association for Computing Machinery, New York, NY, USA (2017), \url{https://doi.org/10.1145/3009837.3009851}

\bibitem{loopy}
Ferdowsifard, K., Barke, S., Peleg, H., Lerner, S., Polikarpova, N.: Loopy: interactive program synthesis with control structures. Proc. ACM Program. Lang.  \textbf{5}(OOPSLA) (Oct 2021). \doi{10.1145/3485530}

\bibitem{snippy}
Ferdowsifard, K., Ordookhanians, A., Peleg, H., Lerner, S., Polikarpova, N.: Small-step live programming by example. In: Proceedings of the 33rd Annual ACM Symposium on User Interface Software and Technology. p. 614–626. UIST '20, Association for Computing Machinery, New York, NY, USA (2020), \url{https://doi.org/10.1145/3379337.3415869}

\bibitem{syndstrans}
Feser, J.K., Chaudhuri, S., Dillig, I.: Synthesizing data structure transformations from input-output examples. SIGPLAN Not.  \textbf{50}(6),  229–239 (Jun 2015). \doi{10.1145/2813885.2737977}

\bibitem{codehint}
Galenson, J., Reames, P., Bodik, R., Hartmann, B., Sen, K.: Codehint: dynamic and interactive synthesis of code snippets. In: Proceedings of the 36th International Conference on Software Engineering. p. 653–663. ICSE 2014, Association for Computing Machinery, New York, NY, USA (2014), \url{https://doi.org/10.1145/2568225.2568250}

\bibitem{stringio}
Gulwani, S.: Automating string processing in spreadsheets using input-output examples. SIGPLAN Not.  \textbf{46}(1),  317–330 (Jan 2011). \doi{10.1145/1925844.1926423}

\bibitem{10.1145/1993316.1993506}
Gulwani, S., Jha, S., Tiwari, A., Venkatesan, R.: Synthesis of loop-free programs. SIGPLAN Not.  \textbf{46}(6),  62–73 (Jun 2011). \doi{10.1145/1993316.1993506}

\bibitem{loopfree}
Gulwani, S., Jha, S., Tiwari, A., Venkatesan, R.: Synthesis of loop-free programs. In: Proceedings of the 32nd ACM SIGPLAN Conference on Programming Language Design and Implementation. p. 62–73. PLDI '11, Association for Computing Machinery, New York, NY, USA (2011). \doi{10.1145/1993498.1993506}, \url{https://doi.org/10.1145/1993498.1993506}

\bibitem{mimic}
Heule, S., Sridharan, M., Chandra, S.: Mimic: computing models for opaque code. In: Proceedings of the 2015 10th Joint Meeting on Foundations of Software Engineering. p. 710–720. ESEC/FSE 2015, Association for Computing Machinery, New York, NY, USA (2015). \doi{10.1145/2786805.2786875}

\bibitem{jain2022jigsaw}
Jain, N., Vaidyanath, S., Iyer, A., Natarajan, N., Parthasarathy, S., Rajamani, S., Sharma, R.: Jigsaw: Large language models meet program synthesis. In: Proceedings of the 44th International Conference on Software Engineering. pp. 1219--1231 (2022)

\bibitem{oguide}
Jha, S., Gulwani, S., Seshia, S.A., Tiwari, A.: Oracle-guided component-based program synthesis. In: Proceedings of the 32nd ACM/IEEE International Conference on Software Engineering - Volume 1. p. 215–224. ICSE '10, Association for Computing Machinery, New York, NY, USA (2010). \doi{10.1145/1806799.1806833}

\bibitem{jiang2024generating}
Jiang, S., Zhu, C., Khurshid, S.: Generating executable oracles to check conformance of client code to requirements of jdk javadocs using llms. arXiv preprint arXiv:2411.01789  (2024)

\bibitem{lamothe2021systematic}
Lamothe, M., Gu\'{e}h\'{e}neuc, Y.G., Shang, W.: A systematic review of api evolution literature. ACM Comput. Surv.  \textbf{54}(8) (Oct 2021). \doi{10.1145/3470133}, \url{https://doi.org/10.1145/3470133}

\bibitem{li2023cctest}
Li, Z., Wang, C., Liu, Z., Wang, H., Chen, D., Wang, S., Gao, C.: Cctest: Testing and repairing code completion systems. In: 2023 IEEE/ACM 45th International Conference on Software Engineering (ICSE). pp. 1238--1250. IEEE (2023)

\bibitem{liu2024llm}
Liu, K., Liu, Y., Chen, Z., Zhang, J.M., Han, Y., Ma, Y., Li, G., Huang, G.: Llm-powered test case generation for detecting tricky bugs. arXiv preprint arXiv:2404.10304  (2024)

\bibitem{reuseableAPI}
Mai, Y., Gao, Z., Hu, X., Bao, L., Liu, Y., Sun, J.: Are human rules necessary? generating reusable apis with cot reasoning and in-context learning. Proc. ACM Softw. Eng.  \textbf{1}(FSE) (Jul 2024), \url{https://doi.org/10.1145/3660811}

\bibitem{jungloid}
Mandelin, D., Xu, L., Bod\'{\i}k, R., Kimelman, D.: Jungloid mining: helping to navigate the api jungle. In: Proceedings of the 2005 ACM SIGPLAN Conference on Programming Language Design and Implementation. p. 48–61. PLDI '05, Association for Computing Machinery, New York, NY, USA (2005). \doi{10.1145/1065010.1065018}

\bibitem{icmlpbe}
Menon, A.K., Tamuz, O., Gulwani, S., Lampson, B., Kalai, A.T.: A machine learning framework for programming by example. In: Proceedings of the 30th International Conference on International Conference on Machine Learning - Volume 28. p. I–187–I–195. ICML'13, JMLR.org (2013)

\bibitem{nam2024using}
Nam, D., Macvean, A., Hellendoorn, V., Vasilescu, B., Myers, B.: Using an llm to help with code understanding. In: Proceedings of the IEEE/ACM 46th International Conference on Software Engineering. pp. 1--13 (2024)

\bibitem{testdriven}
Perelman, D., Gulwani, S., Grossman, D., Provost, P.: Test-driven synthesis. In: Proceedings of the 35th ACM SIGPLAN Conference on Programming Language Design and Implementation. p. 408–418. PLDI '14, Association for Computing Machinery, New York, NY, USA (2014). \doi{10.1145/2594291.2594297}, \url{https://doi.org/10.1145/2594291.2594297}

\bibitem{schaeffer2024emergent}
Schaeffer, R., Miranda, B., Koyejo, S.: Are emergent abilities of large language models a mirage? Advances in Neural Information Processing Systems  \textbf{36} (2024)

\bibitem{tfcoder}
Shi, K., Bieber, D., Singh, R.: Tf-coder: Program synthesis for tensor manipulations. ACM Trans. Program. Lang. Syst.  \textbf{44}(2) (May 2022), \url{https://doi.org/10.1145/3517034}

\bibitem{frangel}
Shi, K., Steinhardt, J., Liang, P.: Frangel: component-based synthesis with control structures. Proc. ACM Program. Lang.  \textbf{3}(POPL) (Jan 2019). \doi{10.1145/3290386}

\bibitem{spiess2024quality}
Spiess, C., Gros, D., Pai, K.S., Pradel, M., Rabin, M.R.I., Jha, S., Devanbu, P., Ahmed, T.: Quality and trust in llm-generated code. arXiv preprint arXiv:2402.02047  (2024)

\bibitem{proveri}
Srivastava, S., Gulwani, S., Foster, J.S.: From program verification to program synthesis. SIGPLAN Not.  \textbf{45}(1),  313–326 (Jan 2010). \doi{10.1145/1707801.1706337}

\bibitem{transit}
Udupa, A., Raghavan, A., Deshmukh, J.V., Mador-Haim, S., Martin, M.M., Alur, R.: Transit: specifying protocols with concolic snippets. In: Proceedings of the 34th ACM SIGPLAN Conference on Programming Language Design and Implementation. p. 287–296. PLDI '13, Association for Computing Machinery, New York, NY, USA (2013). \doi{10.1145/2491956.2462174}, \url{https://doi.org/10.1145/2491956.2462174}

\bibitem{ugare2024improving}
Ugare, S., Suresh, T., Kang, H., Misailovic, S., Singh, G.: Improving llm code generation with grammar augmentation. arXiv preprint arXiv:2403.01632  (2024)

\bibitem{wang2023large}
Wang, X., Zhu, W., Wang, W.Y.: Large language models are implicitly topic models: Explaining and finding good demonstrations for in-context learning. arXiv preprint arXiv:2301.11916 p.~3 (2023)

\bibitem{wei2022chain}
Wei, J., Wang, X., Schuurmans, D., Bosma, M., Xia, F., Chi, E., Le, Q.V., Zhou, D., et~al.: Chain-of-thought prompting elicits reasoning in large language models. Advances in neural information processing systems  \textbf{35},  24824--24837 (2022)

\bibitem{xia2024fuzz4all}
Xia, C.S., Paltenghi, M., Le~Tian, J., Pradel, M., Zhang, L.: Fuzz4all: Universal fuzzing with large language models. In: Proceedings of the IEEE/ACM 46th International Conference on Software Engineering. pp. 1--13 (2024)

\bibitem{Edsynth}
Yang, Z., Hua, J., Wang, K., Khurshid, S.: Edsynth: Synthesizing api sequences with conditionals and loops. In: 2018 IEEE 11th International Conference on Software Testing, Verification and Validation (ICST). pp. 161--171 (2018). \doi{10.1109/ICST.2018.00025}

\end{thebibliography}

\appendix
\section{Prompt}
\begin{figure}
	\centering
\begin{lstlisting}[style = prompt]
<Role>
    You are a software engineer. You will be given with an Java method signature,
    its return type, and a set of test cases as comments. Your task is to implement
    the Java method into a full implementation to pass the test cases.
</Role>
<Instruction>
    Use the following step by step instruction to solve the problem:
        Step 1 - Import Necessary Libraries
            (1) Identify and include any required import statements, such as Java standard libraries (java.util.*), or open-source libraries (e.g., Apache Commons, Guava) to simplify implementation.
            (2) Ensure that all dependencies are properly referenced to facilitate a smooth compilation.
        Step 2 - Define Any Required Helper Classes or Methods
            (1) If the method requires additional data structures or utility functions, define them before implementing the main method.
            (2) Ensure helper classes or methods are designed efficiently to support modularity and reusability.
        Step 3 - Create a Public Class with an Appropriate Name
            (1) Define a class that logically represents the method's functionality (e.g., APIUtility, StringProcessor, MathHelper).
            (2) Ensure the class is structured according to Java best practices.
        Step 4 - Understand the Problem Statement and Requirements
            (1) Analyze the method signature, return type, and test cases to infer the expected behavior.
            (2) Identify edge cases that may not be explicitly covered by the test cases but are crucial for correctness (e.g., null inputs, empty lists, boundary values).
        Step 5 - Implement the Method to Pass All Test Cases
            (1) Write a clear and efficient implementation of the method to meet the expected input-output requirements.
            (2) Optimize for performance and maintainability while ensuring correctness.
        Step 6 - Validate Edge Cases and Complete Implementation
            (1) Consider additional edge cases beyond the provided test cases (e.g., large inputs, invalid values).
            (2) Refactor the code if necessary to handle all identified cases.
        Step 7 - Write Unit Tests for Each Provided Test Case
            (1) Create a test method for each given test case using JUnit (or another testing framework).
            (2) Ensure test methods cover normal, edge, and erroneous inputs.
        Step 8 - Provide a Main Method for Execution and Validation
            (1) Implement a main method to run and validate the test cases.
            (2) Print test results to confirm correctness and ensure all test cases pass.
            (3) Verify that the code compiles and runs without errors.
</Instruction>
<examples> 
    Here is one example:
    <method>
        List<Integer> GetRange(int start, int end)
    </method>
    <TestCases>
        1. start = 10, end = 9 -> output: []
        2. start = 10, end = 10 -> output: []
        3. start = 10, end = 11 -> output: [10]
        4. start = 10, end = 12 -> output: [10, 11]
        5. start = -2, end = 2 -> output: [-2, -1, 0, 1]
    </TestCases>
    <output>
        List<Integer> GetRange(int start, int end) {
            ArrayList<Integer> range = new ArrayList<Integer>();
            for (int i=0; i+start < end; i++)
                range.add(Integer.valueOf(i+start));
            return range
        }
    </output>
</examples>
Now, give you the following method signature and test cases:
<method>
     double ellipseArea(Ellipse2D ellipse)
</method>
<TestCases>
    1.  Ellipse2D.Double(12.3, -45.6, 7.8, 9) -> 3.9 * 4.5 * Math.PI
    2.  Ellipse2D.Double(12.3, -45.6, 7.8, 2) -> 3.9 * Math.PI
    3.  Ellipse2D.Double(12.3, -45.6, 2, 7.8) -> 3.9 * Math.PI
    4.  Ellipse2D.Double(12.3, -45.6, 2, 2) -> Math.PI
</TestCases>

Please output the results. Only output complete code.
 \end{lstlisting}
 \caption{An example of our prompt input to LLMs}
		\label{fig:figure4}
\end{figure}

\end{document}